# Gland Segmentation in Histopathological Images by Deep Neural Network

Safiye Rezaei, Ali Emami, Nader Karimi, Shadrokh Samavi
Isfahan University of Technology
Isfahan, 84156-83111 Iran.

*Abstract*—Histology method is vital in the diagnosis and prognosis of cancers and many other diseases. For the analysis of histopathological images, we need to detect and segment all gland structures. These images are very challenging, and the task of segmentation is even challenging for specialists. Segmentation of glands determines the grade of cancer such as colon, breast, and prostate. Given that deep neural networks have achieved high performance in medical images, we propose a method based on the LinkNet network for gland segmentation. We found the effects of using different loss functions. By using Warwick-Qu dataset, which contains two test sets and one train set, we show that our approach is comparable to state-of-the-art methods. Finally, it is shown that enhancing the gland edges and the use of hematoxylin components can improve the performance of the proposed model.

## I. Introduction

One of the primary methods for cancer prognosis and diagnosis is tissue biopsy. A tissue sample is placed on a glass slide and is stained for inspection. Then a pathologist examines the sample using a microscope to see if cancer is present. Histopathology is a microscopic examination of the samples to determine cancer by tissue pathologists [1].

Since tumors have a high degree of cellular integrity, the classification and segmentation of tissue cells in histopathology images are essential. This is one of the cheapest morphological methods that pathologists can use to study tissue samples on a large scale. Furthermore, collecting histopathology images is quick and has little risk to patients. However, the analysis of the collected slides is difficult and time-consuming due to the complex nature of images and human recognition errors.

Although histopathology has about one hundred years of history, pathologists estimate that they will be used for the next fifty years due to the efficacy of the method. Detection of histopathological structures like glands, lymphocytes, and cancerous nuclei, is a prerequisite for the diagnosis and prognosis of cancer in histopathology images. The morphological appearance of these structures contains essential information. For instance, the shape of glands in malignant cases implies the type of deficiency. In colon cancer, it has irregular shape while in benign case, it has a regular circular shape. Another reason for the segmentation of histopathological images is cell counting, which can help to diagnose some types of cancers [1].

Gland segmentation in histopathology images is a challenging task, since some glands may be interconnected or have irregular shapes. On the other hand, manual segmentation is time-consuming and results in a less accurate diagnosis. Hence, the automatic segmentation of glands is still an open research problem in the community.

Researchers propose different methods for gland segmentation. For example, Paul et al. [2] used structural information for gland segmentation. They proposed an edge-preserving filter for keeping the boundary of glands, then segmented the glands using informative morphological scale space. This method is faster than segmentation based on neural networks [2]. Furthermore, they showed that the red channel in RGB histopathology images has rich information for segmentation. However, the experimental results are not comparable to neural network-based approaches.

Many research works have proposed to segment the glands by using neural networks. Zhang et al. [3] proposed to decompose the original ground truth based on the shape structures and applied a K-to-1 deep network to solve the segmentation problem using multiple segmentation subproblems. Recently some works use handcrafted features and deep features. Monivannan et al. [4] compute some hand-crafted features such as multi-resolution local binary patterns (LBP) and SIFT for sliding windows of the image, which are then clustered into 30 groups by K-means and classified by SVM. In a subsequent work [5], authors add deep features to achieve improved results. Deep features are extracted from a pre-trained fully convolutional neural network.

Chen et al. [6] propose to segment glands and their contours, which are combined to compute the final segmentation map. In [7] Graham et al. offer a network based on a minimal information loss unit to preserve missing details during downsampling operation by incorporating a down-sampled original image into each residual unit. Furthermore, they segmented the gland lumen to improve their results. They also perform statistical analysis on network outputs by feeding various transformed images as network input.

Histopathology images are generally stained with Hematoxylin and Eosin technique, which provide more discriminative color information [8]. For a better representation of the color image information, stain decomposition is used. Rezaei et al. [9] propose using handcrafted features and Hematoxylin components as network input. They show that the red channel and the Hematoxylin component have more information for segmentation. Furthermore, they use two loss functions for network training to improve the model results. Wang et al. [10] classify histopathological images by using stain decomposition of H&E components as inputs to a bilinear CNN.

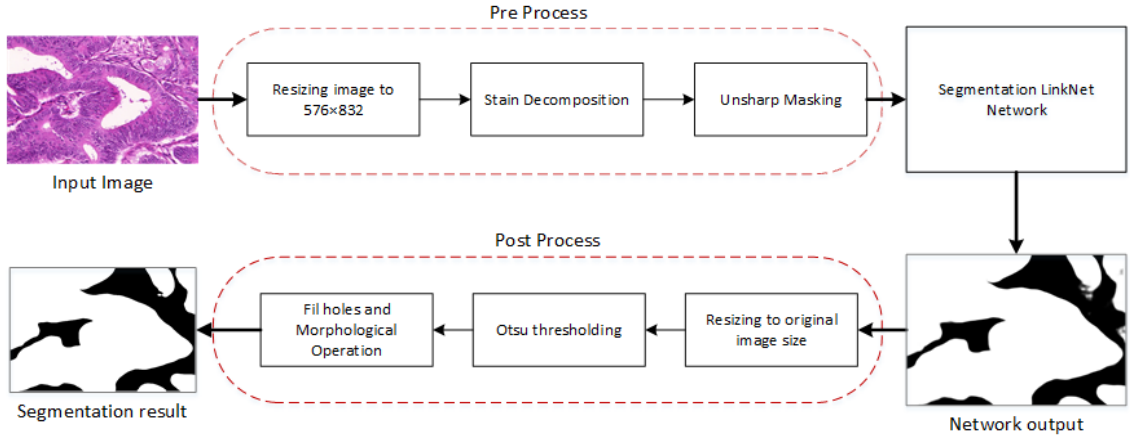

Fig. 1. Structure of the proposed model

Kadkhodaei et al. [11] use unsharp masking to enhance the contrast of MRI image channels to emphasize the edges in the preprocessing stage. This helps super-voxel segmentation and increases the quality of texture and saliency features for classification.

In this paper, we propose a deep neural network with multiple loss functions based on LinkNet structure [12] for gland segmentation. LinkNet structure is structurally similar to UNet [13]. However, LinkNet has fewer trainable parameters, which makes it faster. We demonstrate that our proposed method has comparable performance to the state-of-the-art. The technical details of the proposed system are provided in section 2, and our experimental results are presented in section 3.

## II. PROPOSED METHOD

Our proposed method consists of three units as shown in Fig.1. In the first unit, preprocessing operations such as resizing and unsharp-masking of the input images, as well as extraction of hematoxylin components, are performed to prepare the input data for the network. The second unit is a modified version of LinkNet used for gland segmentation (Cf. Fig.3). The post-processing module performs morphological operation and Otsu thresholding to generate the final segmentation map. The technical details of each unit are described in the following sections.

### A. preprocessing

Since images of the Warwick-QU [14] dataset have different sizes, we need to resize them to the same size. Then unsharp-masking is applied for deblurring and making stronger edges. According to [11] this technique improves the quality of image segmentation. We examine the effectiveness of this technique in the experimental results. Sample of histopathology image, hematoxylin component of it, and applying unsharp masking of it, shown in Fig 2.

Furthermore, Hematoxylin component is extracted from histopathological RGB images similar to [8, 9]. Stain decomposition is achieved based on an orthonormal transformation of RGB channels. It is shown that the boundaries of glands in the hematoxylin component are more visible [9].

### B. Network structure

The proposed segmentation network is a modified version of LinkNet network, in which we change and add loss functions to specific points of the network to improve the segmentation accuracy. LinkNet is an autoencoder model, including encoder and decoder blocks as shown in Fig.2. It can be considered as a light version of UNet with one-third parameters (11 million vs. 33 million parameters), while its performance is comparable to UNet. This increases the speed of training and testing for image segmentation.

As demonstrated in Fig.3, in the proposed network, we define a new output driven by the final layers of the network and train it to provide a coarse-scale segmentation map of the glands. Hence, we apply a separate loss function for training this output. Consequently, we have two loss functions for training the network: $L_i$ for internal output and $L_o$ for final output. The training loss function used for training is a weighted sum of $L_i$ and $L_o$, as shown in equation (1). As pointed earlier, the initial output performs a primary segmentation task on a coarse scale, and the final output builds up a detailed segmentation map with full resolution. Intuitively, we give a higher weight to the initial loss function.

$$L_{final} = 2 \times L_i + L_o \qquad (1)$$

Let's assume G is the ground truth, and O is the network output. Then our loss functions are defined by the combination of binary cross-entropy (BCE) (Eq.2), dice (Eq.3), and accuracy (Eq.4). Cross entropy decreases as the outputs get closer to the ground truth. Binary cross-entropy and accuracy are pixel-wise metrics, while dice is a holistic metric, which represents the total similarity between the estimated segmentation map and ground-truth.

$$BCE(G,O) = -[G \times log(O) + (1-G) \times log(1-O)] \qquad (2)$$

$$D(G,O) = \frac{2 \times |G \cap O| + S}{|G| + |O| + S} \qquad (3)$$

$$Accuracy = \frac{(TP+TN)}{(TP+TN+FP+FN)} \qquad (4)$$

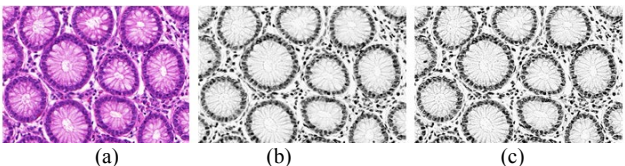

(a)　　　　(b)　　　　(c)

Fig. 2. (a) sample of histopathology image, (b) hematoxylin component of it, (c) after applying unsharp masking

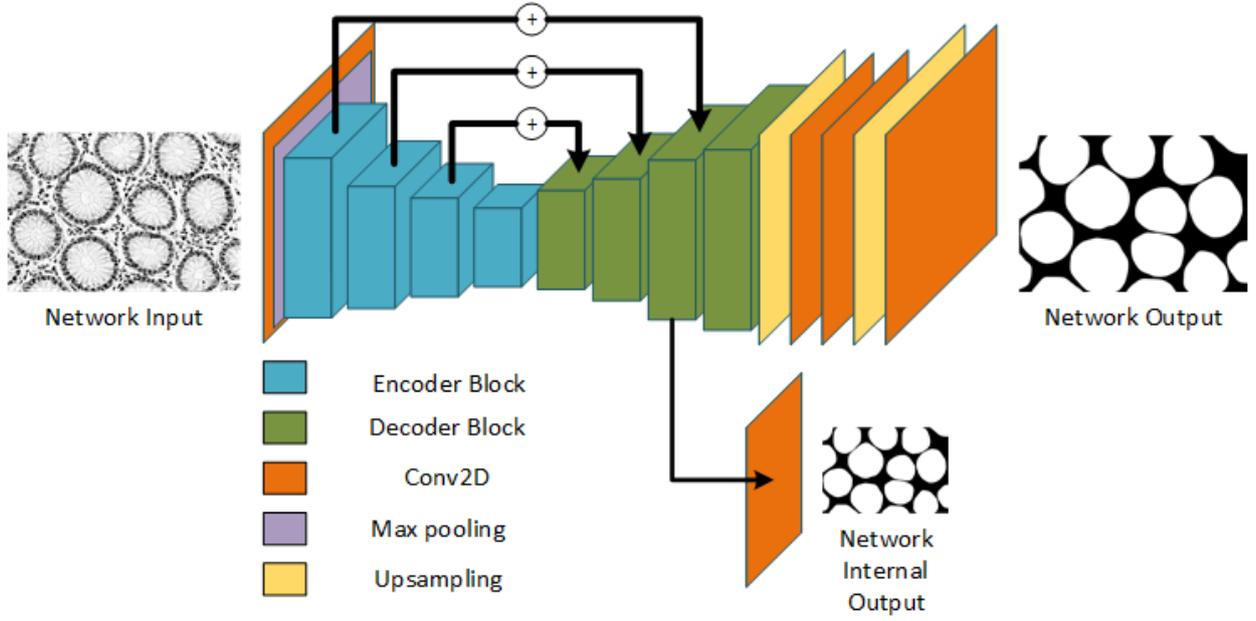

Fig. 3. The Network Structure

For $L_i$ and $L_o$, we test three loss functions $L_1$ to $L_3$ shown by equations (Eq.5 - Eq.7). By combining pixel-wise and holistic metrics, the network can be trained more efficiently. As shown in the next section, experimental results imply that the third loss function $L_3$ facilitates the best training and achieves the best results. This is in-line with our expectation, as the proposed loss function considers all the three metrics, with an emphasis on the pixel-wise metrics.

$$L1 = BCE(G,O) - e^{(1+Dice(G,O))} \quad (5)$$

$$L2 = BCE(G,O) - e^{(1+Dice(G,O))} - Accuracy \quad (6)$$

$$L3 = 2 \times BCE(G,O) - e^{(1+Dice(G,O))} - Accuracy \quad (7)$$

C. *Post-process*

We resize the output segmentation map to the original size of the image to compare the ground truth and segmentation result in the testing phase. One of the best methods for automatic thresholding and generating a binary map is the Otsu method. Therefore, Otsu thresholding is applied to convert the probability map into binary. The generated binary map at this stage has some noises and holes in the glands. Morphological operations are performed on the binary map for denoising and filling the holes towards producing a solid segmentation map.

### III. EXPERIMENTAL RESULTS

We implemented our proposed method by the TensorFlow library in python on a desktop system with 32GB Ram and NVIDIA GeForce Titan X GPU. Training time on this system is about 7 hours for 100 epochs. We use public Warwick-QU dataset[1] for training and evaluation. This dataset contains 85 color images for training and two sections for testing. Section A has 60 images, and section B has 20 images. Since we do not have enough training images, the training images are augmented to generate 4420 images. There is no available LinkNet model for medical applications. Thus we do not use any pre-trained network.

A. *Data Preparation*

We resize all of 85 training images to size $(832 \times 576)$, then we augment these images by rotation and flipping them in horizontal and vertical directions, then we extract four same-size cropped images from each image. Finally, we create 4420 images.

B. *Evaluation Metrics*

Following prior works [6], which reported object-level Dice, F1 score, and Hausdorff distance as evaluation metrics, we report these metrics for our results, too. Object-level dice, Hausdorff distance, and object-level Hausdorff are defined by equations (Eq.7, Eq.8 & Eq.9).

$$D_{object}(G,O) = \frac{1}{2}\left(\sum_{i=1}^{n_I} w_i D(G_i,O_i) + \sum_{j=1}^{n_G} \widetilde{w_j} D(\widetilde{G_j},\widetilde{O_j})\right) \quad (7)$$

In equation (7) and (8) $O_i$ is the $i^{th}$ gland in the output of network and $G_i$ is the ground truth gland that has the maximum overlap with $O_i$. $\widetilde{G_j}$ is the $j^{th}$ ground truth gland and $\widetilde{O_j}$ is the segmented gland with maximum overlap with $\widetilde{G_j}$. The weights $w_i$ and $\widetilde{w_j}$ are defined as $w_i = \frac{|O_i|}{\sum_{k=1}^{n_I}|O_k|}$ and $\widetilde{w_j} = \frac{|\widetilde{G_j}|}{\sum_{k=1}^{n_G}|\widetilde{G_k}|}$.

$$H(G,O) = max\left\{\sup_{x \in G}\inf_{y \in O}\|x-y\|, \sup_{y \in O}\inf_{x \in G}\|x-y\|\right\} \quad (8)$$

$$H_{object}(G,O) = \frac{1}{2}\left(\sum_{i=1}^{n_O} w_i H(G_i,O_i) + \sum_{j=1}^{n_G} \widetilde{w_j} H(\widetilde{G_j},\widetilde{O_j})\right) \quad (9)$$

If the intersection between *O* and *G* *is* greater than fifty percent, the F1 score can be defined as below:

---

[1] http://www2.warwick.ac.uk/fac/sci/dcs/research/combi/research/bic/ glascontest/

$$F1 = \frac{2PR}{P+R}, P = \frac{TP}{TP+FP}, R = \frac{TP}{TP+FN} \qquad (10)$$

*C. Evaluation Results*

For the A and B test sections, we test our trained model and show the results in Table 1. Also, the results of a state-of-the-art model and our prior work are compared against the proposed model.

Table 1. Segmentation results for the A test-set

| Method | Object Dice | F1 Score | Hausdorff |
|---|---|---|---|
| Hemotoxillin component and Loss Function L1 | 0.869 | 0.812 | 56.84 |
| Hemotoxillin component and Loss Function L2 | 0.858 | 0.823 | 63.9 |
| Hemotoxillin component and Loss Function L3 | 0.874 | 0.846 | 55.13 |
| Hemotoxillin component, Unsharp masking and Loss Function L1 | 0.849 | 0.817 | 66.13 |
| Hemotoxillin component, Unsharp masking and Loss Function L2 | 0.872 | 0.832 | 55.51 |
| Hemotoxillin component, Unsharp masking and Loss Function L3 | 0.823 | 0.773 | 79.1 |
| Rezaei et al. [9] | 0.867 | 0.83 | 56.641 |
| CUMedVision1 [6] | 0.867 | 0.868 | 74.596 |
| CUMedVision2 [6] | 0.897 | 0.912 | 45.418 |

As shown in Table 1, when the hematoxylin component is chosen for network input, and L3 is used as the network loss function, the best result is achieved by our model. On the other hand, when Unsharp masking is applied to the hematoxylin component and L2 loss is used, the second-best result is obtained.

Table 2. Segmentation results for the B test-set

| Method | Object Dice | F1 Score | Hausdorff |
|---|---|---|---|
| Hemotoxillin component and Loss Function L1 | 0.82 | 0.73 | 109.72 |
| Hemotoxillin component and Loss Function L2 | 0.798 | 0.695 | 129.77 |
| Hemotoxillin component and Loss Function L3 | 0.816 | 0.739 | 118.95 |
| Hemotoxillin component, Unsharp masking and Loss Function L1 | 0.816 | 0.731 | 123.66 |
| Hemotoxillin component, Unsharp masking and Loss Function L2 | 0.817 | 0.752 | 111.17 |
| Hemotoxillin component, Unsharp masking and Loss Function L3 | 0.765 | 0.72 | 161.85 |
| Rezaei et al. [9] | 0.822 | 0.75 | 108.208 |
| CUMedVision1 [6] | 0.8 | 0.769 | 153.646 |
| CUMedVision2 [6] | 0.781 | 0.716 | 160.347 |

The B test-set contains more challenging images than the A test-set, which makes the segmentation more difficult. The segmentation results of section B shown in Table 2, demonstrate more closely competing results. In the proposed model hematoxylin component with L1 loss without unsharp masking, and hematoxylin component with Unsharp masking and L2 loss provide the best results.

IV. CONCLUSION

In this paper, we used modified LinkNet with two loss functions and compared the effect of using different inputs and loss functions in results. Loss functions are a combination of accuracy, binary cross-entropy, and dice metrics. We also showed that the use of hematoxylin component of a histopathology image could reveal more information about the glands and could facilitate a more accurate segmentation. Our best results are comparable to state-of-the-art methods.